\documentclass[floatfix,a4paper,tightenlines,amsmath,amssymb,nofootinbib,showpacs,notitlepage,showkeys,prd,twocolumn,superscriptaddress,10pt,aps]{revtex4-1}

\usepackage[utf8]{inputenc}
\usepackage{amsmath}
\usepackage{amssymb}
\usepackage{graphicx}
\usepackage{hyperref}
\usepackage{floatrow}
\usepackage{subcaption}

\newcommand{\eref}[1]{Eq. (\ref{#1})}
\newcommand{\fref}[1]{Fig. \ref{#1}}

\newcommand{\myhfill}{\hfill\textcolor{white}{.}}

\allowdisplaybreaks 

\begin{document}

\title{Spectral dimension as a tool for analyzing nonpertubative propagators}

\author{Wolfgang~Kern}
\email{wolfgangjohannkern@gmail.com}
\affiliation{Institute of Physics, University of Graz, NAWI Graz, Universit\"atsplatz 5, 8010 Graz, Austria}
\author{Markus Q.~Huber}
\email{markus.huber@physik.jlug.de}
\affiliation{Institut f\"ur Theoretische Physik, Justus-Liebig--Universit\"at Giessen, 35392 Giessen, Germany}
\author{Reinhard~Alkofer}
\email{reinhard.alkofer@uni-graz.at}
\affiliation{Institute of Physics, University of Graz, NAWI Graz, Universit\"atsplatz 5, 8010 Graz, Austria}

\date{\today}

\begin{abstract}
We derive general properties of the scale-dependent effective spectral dimensions of nonperturbative gauge boson propagators as they appear as solutions from different methods in Yang-Mills theories.
In the ultraviolet and for short timescales the anomalous dimensions of the propagators lead to a slight decrease of the spectral dimension as compared to the one of a free propagator.
Lowering the momentum scale, the spectral dimension decreases further.
The class of propagators which display a maximum at Euclidean momenta, and thus violate positivity, always approaches a spectral dimension of one for large times. 
We also show that the longest time intervals are {\em not} related to the deep infrared but to the momentum scale defined by the position of the maximum.
\end{abstract}



\maketitle

\section{Introduction}
\label{sec:introduction}

The interest in the fully renormalized gluon propagator dates back into the late seventies of the last century, see, {\it e.g.}, 
Refs.\ \cite{Pagels:1976kq,Delbourgo:1978jw,Oehme:1979ai}, and the behavior of the gluon propagator in the strongly interacting domain of quantum chromodynamics (QCD) is in the focus of research until today, 
see, {\it e.g.},  \cite{Alkofer:2000wg,Maas:2011se,Aguilar:2015bud,Huber:2018ned} and references therein.
The question whether the gluon is not only confined but also confining was raised already in the first investigations of this topic \cite{Pagels:1976kq}, and the thus motivated assumption of an infrared singular gluon propagator received some support from the observation that it might be responsible for the area law in the Wilson loop \cite{West:1982bt} and thus quark confinement.
Nowadays, however, it is generally accepted that in the Landau gauge the gluon propagator is at small momenta suppressed as compared to a free propagator, for recent reviews see, {\it e.g.},  \cite{Maas:2011se,Aguilar:2015bud,Huber:2018ned}.\footnote{
This suppression exists for the gauge boson propagator in Yang-Mills theory, {\it i.e.}, the quenched gluon propagator, and 
for the unquenched gluon propagator as long as the number of light quarks is small. For a large number of flavors such that one enters the conformal window a change in the behavior of the gluon propagator is expected \cite{Hopfer:2014zna}.}
Although it is evident that such a suppression implies positivity violation \cite{Oehme:1979ai,Oehme:1994pv}, the relation to gluon confinement stays elusive, and the understanding of the analytic structure of the gluon propagator, despite some progress in this direction, e.g., \cite{Alkofer:2003jj,Strauss:2012dg,Lowdon:2017uqe,Lowdon:2018mbn,Cyrol:2018xeq,Binosi:2019ecz}, is still unsatisfactory.

Therefore the question arises whether the gluon propagator can be analyzed with the help of other tools. Herein, we are going to employ in this respect the concept of a spectral dimension. Introduced in Ref.\ \cite{Ambjorn:2005db} in the context of causal dynamical triangulation studies in the research area of quantum gravity, it has found wide-spread use in analyzing spacetimes resulting from different approaches to quantum gravity, see, {\it e.g.},  \cite{Carlip:2017eud,Carlip:2019onx} for a recent compilation. Hereby, different attempts toward quantum gravity typically see a dimensional reduction when going from the infrared to the ultraviolet. Technically, in some of these studies this is achieved by studying the spectral dimension $D_S$ via the diffusion of a hypothetical scalar particle,  and instead of taking the limit of infinite diffusion times as in the mathematical definition of the spectral dimension one studies a generalized effective spectral dimension $D_S(T)$ as a function of the diffusion time $T$ \cite{Lauscher:2005qz}. The quantity $D_S(T)$ is closely related to the two-point correlation function of the diffusing particle. Therefore, given the propagators as resulting from a certain approach to quantum gravity, one can perform investigations how the effective dimension of spacetime changes, see \cite{Alkofer:2014raa} for an example of such a study. 

In the context of the here presented results it is important to note that the spectral dimension can be interpreted as Haussdorff dimension of momentum space \cite{Amelino-Camelia:2013gna,Alkofer:2014raa,Alkofer:2018zze}. This makes evident that the concept of a spectral dimension is not restricted to study properties of a spacetime by using propagators but one can also employ it to analyze propagators with respect to their nontrivial content due to interactions. Local interactions which respect unitarity will always make the spectrum as measured by the propagator ``thinner'', {\it i.e.}, we expect in a local quantum field theory, and thus in Yang-Mills theory as well as QCD, a reduction of the spectral dimension as compared to the topological dimension of the underlying spacetime due to interactions. This reduction will be larger for stronger interactions, and the hypothetical diffusion time will allow us to draw conclusions of the scale-dependent impact of interactions onto the propagator. As we will argue below, one important result of our investigation is that for the class of propagators the gluon propagator belongs to, a simple mapping of large diffusion times to small momenta is not correct. The maximum displayed by such propagators is then not only indicative of positivity violation but also responsible for breaking the ``rule-of-thumb'' that the longest time intervals and largest distances relate to the smallest momenta.

This makes also plain that there is another technical distinction to quantum gravity studies which are generically interested in sub-Planckian and therefore the shortest distances. These are related, of course, to the shortest diffusion times. 
As Yang-Mills theories are asymptotically free the interactions are small for large momenta and short distances, and the expected relation to short diffusion times is verified. In two and three dimensions we find accordingly the topological dimension as value for the short-time spectral dimension, while in four dimensions we find a reduction which we can relate to the logarithmic corrections encoded in the anomalous dimension of the gluon propagator. The most interesting momentum regime for our study is, of course, the strongly interacting one. Somewhat surprisingly it turns out that the largest diffusion times are {\em not} related to the deep-infrared behavior of the propagator.
Thus, the spectral dimension highlights different features than other integral transformations, as, e.g., the Fourier transformation.

In the next section we discuss the effective spectral dimension and some of its properties.
Sec.~\ref{sec:input} describes the input for which we calculate in Sec.~\ref{sec:results} the spectral dimension as a function
of the diffusion time.
We summarize and conclude in Sec.~\ref{sec:conclusions}.
Some details of the input and proofs are deferred to appendices.

\section{General considerations for the spectral dimension}
\label{sec:spec_dim}

\subsection{The spectral dimension}

We calculate the spectral dimension $D_S(T)$ from the return probability as explained in Ref.~\cite{Alkofer:2014raa}.
Diffusion is governed by the heat kernel $K(x,x',T)$ which describes quantitatively the diffusion from a start position $x$ to a final position $x'$ in some diffusion time $T$.
The heat kernel satisfies the heat kernel equation 
\begin{align}
\label{eq:K}
 \left(\frac{\partial}{\partial T}-\Delta_x\right)K(x,x',T)=0
\end{align}
with
\begin{align}
K(x,x',0)=\delta(x-x').
\end{align}
In this equation, $\Delta_x$ is the Laplacian operator.
We can solve the equation by Fourier transformation and obtain
\begin{align}
 K(x,x',T)= \int \frac{d^d p}{(2 \pi)^d}e^{ip(x-x')}e^{-p^2 T}.
\end{align}
Generally, one can identify $K(x,x',T)$ with matrix elements of the operator $e^{\Delta_x T}$.
The return probability $\mathcal{P}(T)$ is related to $K(x,x',T)$ via 
\begin{align}
 \mathcal{P}(T)=V^{-1}\int d^dx \sqrt{g(x)}K(x,x,T) = V^{-1} \text{Tr}\left( e^{\Delta_x T}\right),
\end{align}
viz., the average over the heat kernel gives the return probability.
The total volume is given by $V=\int d^dx \sqrt{g(x)}$ where $g(x)$ is the metric.
In flat space the return probability for a free propagators is
\begin{align}
\label{eq:PT_free_part}
\mathcal{P}(T)=(4\pi T)^{-d/2}.
\end{align}
This leads to a spectral dimension, given by
\begin{align}
\label{eq:SpecDimDef}
D_S(T)=-2\dfrac{\partial~ \ln \mathcal{P}(T)}{\partial ~ \ln T}=-2 T\frac{\mathcal{P}'(T)}{\mathcal{P}(T)}.
\end{align}
of $D_S=d$ for any $T$.

Here we will use the Euclidean metric throughout.
Anomalous behavior is then introduced by higher order derivative terms, which correspond to deviations from the classical behavior.
Thus, we need to take into account quantum fluctuations.
Since the Laplace operator $\Delta_x$ corresponds to the classical inverse propagator of a massless particle, we replace it now by the dressed inverse propagator $F$.
Eq.~\ref{eq:K} changes then to
\begin{align}
\left( \partial_T + F(-\Delta_x)\right)K_g(x,x',T)=0.
\end{align}
In an analogous way to above, we obtain the solution for this equation,
\begin{align}
K(x,x',T)=\int \frac{d^dp}{(2 \pi)^d}e^{ip(x-x')}e^{-F(p^2)T},
\end{align}
and for the return probability,
\begin{align}
\label{eq:return_prob}
\mathcal{P}(T)=\int \frac{d^d p}{(2\pi)^d}e^{-F(p^2)T}.
\end{align}

In the following we only consider a flat spacetime.
The spectral dimension of a free massless particle is then equal to the topological dimension of spacetime, see \eref{eq:PT_free_part}.
However, we expect that interactions between particles lower the number of possible energy states which, in turn, lowers the spacetime dimension felt by test particles.
An intuitive way to see this is to transform the integration variable as $k^2=F(p^2)$  \cite{Amelino-Camelia:2013gna,Amelino-Camelia:2013cfa} which leads to the following expression:
\begin{align}
\label{eq:return_prob_k}
 \mathcal{P}(T)=\frac{\Omega_d}{(2\pi)^d}\int dk\,\frac{k\, (F^{-1}(k^2))^\frac{d-2}{2}}{F'(p^2)} e^{-k^2\,T}.
\end{align}
$\Omega_d$ is the angular integral.
This can be interpreted as the return probability of a free propagator in a nontrivial background determined by $F(p^2)$, viz., the quantum fluctuations of the propagator.
A requirement for this transformation is that there is a one-to-one correspondence between $k$ and $p$.

The integral \eref{eq:return_prob} can be interpreted as a nonlinear integral transformation.
A direct consequence of the function being in the exponent is a different relation between $p^2$ and $T$ as compared to typical well-known integral transformations.
For example, low momenta are typically related to large distances via a Fourier transformation.
Here, however, low values of $F(p^2)$ are related to large $T$.
If $F(p^2)$ is monotonic and vanishes at $p^2=0$, small $p^2$ and large $T$ are indeed related.
However, if the inverse propagator vanishes at $p_0\neq 0$, long diffusion times are related to $F(p^2)$ at $p^2\approx p_0^2$.
The properties of the integral transformation discussed above are inherited by the spectral dimension $D_S(T)$ defined in \eref{eq:SpecDimDef}.

\subsection{Asymptotics of the spectral dimension}

We will now discuss some general properties of the spectral dimension before we turn to a full numerical analysis in Sec.~\ref{sec:results}.
We will outline how some specific problems of our input are overcome.
We start the discussion with the simple example of a massive propagator and will explain why using the same method for Yang-Mills theory fails.
We will then turn to two analytic examples that correspond to the asymptotics of the input employed in Sec.~\ref{sec:results}.

For a massive particle the inverse propagator reads
\begin{align}
 F(p^2)=p^2+m^2.
\end{align}
The return probability is\footnote{The prefactors cancel out in the spectral dimension, so they are ignored here.}
\begin{align}
 \mathcal{P}(T)\propto\frac{e^{-m^2\,T}}{2T^2}
\end{align}
from which the spectral dimension is calculated in four dimensions as
\begin{align}
 D_S(T)=4+2\,m^2\,T.
\end{align}
For $m=0$ the spectral dimension of the free massless propagator is obtained.
However, the mass leads to a contribution linear in $T$.
This property remains true for a more general inverse propagator where also a constant term is separated:
\begin{align}
 F(p^2)=F^{(0)}(p^2)+m^2.
\end{align}
It was shown in Ref.~\cite{Alkofer:2014raa} that this leads to a spectral dimension
\begin{align}
 D_S(T)=2\,m^2\,T+D_S^{(0)}(T) 
\end{align}
where $D_S^{(0)}(T)$ is calculated from $F^{(0)}(p^2)$.
Thus, the authors of \cite{Alkofer:2014raa} suggested to remove the trivial linear term by taking $m^2=F(0)$ and considering $D_S^{(0)}(T)$ instead of $D_S(T)$.

However, this procedure runs into problems if the inverse propagator is not monotonic and has a minimum.
Then the subtracted inverse propagator $F(p^2)-F(0)$ is negative in some region which leads to a negative spectral dimension for sufficiently large diffusion times, see Appendix~\ref{sec:neg_spec_dim} for a proof.

In principle, one can calculate the return probability including the linear term and extract the coefficient from a fit.
However, for large $T$ the return probability is suppressed exponentially and the computation of the spectral dimension is numerically not stable.
Another method is to subtract the inverse propagator at the minimum $p_0$:
\begin{align}
 F(p^2)\rightarrow F(p^2)-F(p_0^2).
\end{align}
This method turned out to be the most stable one and is used throughout this work.
However, \eref{eq:return_prob_k} does no longer hold because the propagator is not a bijective function.
This could be circumvented by introducing an IR cutoff at the minimum.
We compared calculations with and without a cutoff at the minimum and only found quantitative differences for intermediate diffusion times.
The asymptotic behavior is not affected, as the minimum itself dominates long diffusion times for both cases.
Thus all calculations shown here were done without a cutoff.

We turn now to some general properties of the spectral dimension.
In Ref.~\cite{Alkofer:2014raa} it was shown that for a polynomial inverse propagator the behavior at short/long diffusion times depends on the highest/lowest exponent of $p^2$.
For the convenience of the reader the proof of this is given in App.~\ref{sec:asymptotics}.
Here we are interested in nonpolynomial inverse propagators.
At short diffusion times the leading behavior is still determined by the highest exponent $N_\text{max}$, which is $1$ here.
Thus, we have
\begin{align}
 \lim_{T\rightarrow 0} D_S(T) = d/N_\text{max}=d,
\end{align}
where $d$ is the dimension of spacetime.
However, in QCD the propagators acquire an anomalous dimension and thus do not show a pure $1/p^2$ behavior.
This will be reflected in the spectral dimension which approaches $4$ very slowly.
Indeed, in simple test models containing a logarithm one can show that the exponent of the logarithm is reflected in the spectral dimension for short diffusion times.
If spectral dimensions $D_{S,1}(T)$ and $D_{S,2}(T)$ with exponents $\gamma_1$ and $\gamma_2$, respectively, are compared, one finds that the ratio $(4-D_{S,1}(T))/(4-D_{S,2}(T))$ is equal to $\gamma_1/\gamma_2$ for small $T$.

As mentioned above, the behavior of the spectral dimension for long diffusion times is related to the behavior of $F(p^2)\rightarrow 0$.
Due to the employed subtraction procedure, this happens at the minimum $p_0^2$.
We can approximate $F(p^2)$ in this region by $a(p^2-p_0^2)^2$, where $a$ is an irrelevant constant.
In that case, the spectral dimension is $1$ for asymptotically large diffusion times.
As shown in App.~\ref{sec:asymptotics}, the spectral dimension for an inverse propagator of the form $(p^2-p_0^2)^c$ with $c>0$ obeys
\begin{align}
\lim_{T\rightarrow \infty}D_S(T)=2/c.
\end{align}

These analytic findings lead to an interesting connection to confinement.
The spectral representation of the propagator $D(p^2)$ reads
\begin{align}\label{eq:KaellenLehmann}
 D(p^2)=\int_{0}^\infty ds \frac{\rho(s)}{p^2+s}.
\end{align}
This form assumes that there are no complex poles.
However, they can be included by taking into account the corresponding residues, see, e.g., \cite{Alkofer:2000wg,Maas:2011se,Huber:2018ned} for more details.
Thus, \eref{eq:KaellenLehmann} encodes the complete analytic structure of the propagator.
In particular, the input in Sec.~\ref{sec:input} was obtained with Euclidean metric, but \eref{eq:KaellenLehmann} applies for arbitrary complex $p^2$.
In general, \eref{eq:KaellenLehmann} holds (in the absence of complex poles), but a necessary condition for a physical particle is that $\rho(s)$ is not negative for any $s$.
Thus, if one can show what $\rho(s)$ is negative somewhere, the particle is removed from the physical state space.
For a propagator with a maximum it is straightforward to show that it violates positivity.
For this we consider the derivative of \eref{eq:KaellenLehmann},
\begin{align}
 \frac{dD(p^2)}{dp^2}=-\int_{0}^\infty ds \frac{\rho(s)}{(p^2+s)^2}.
\end{align}
If the propagator is not monotonic, for example, because it has a maximum, its derivative is zero at this maximum.
This, however, entails, that $\rho(s)$ must be negative in parts of the integration domain such that the integral can vanish.
While violation of positivity is sufficient to remove the gluon from the physical spectrum, it is not a required condition.
Thus, we can only conclude that a gluon propagator with a maximum violates positivity \textit{and} has a spectral dimension of one for long diffusion times.
This is one of the main results of this work.

In Sec.~\ref{sec:results} we numerically confirm the behavior for short and long diffusion times discussed above in all our calculations.

\section{Input}
\label{sec:input}

We summarize the propagators used for the calculations of spectral dimensions in this section.
For easier comparison the propagators are shown in Sec.~\ref{sec:results} together with their spectral dimensions.

\begin{figure*}
	\ffigbox{
		\begin{subfloatrow}
			\ffigbox{\includegraphics[width=0.49\textwidth]{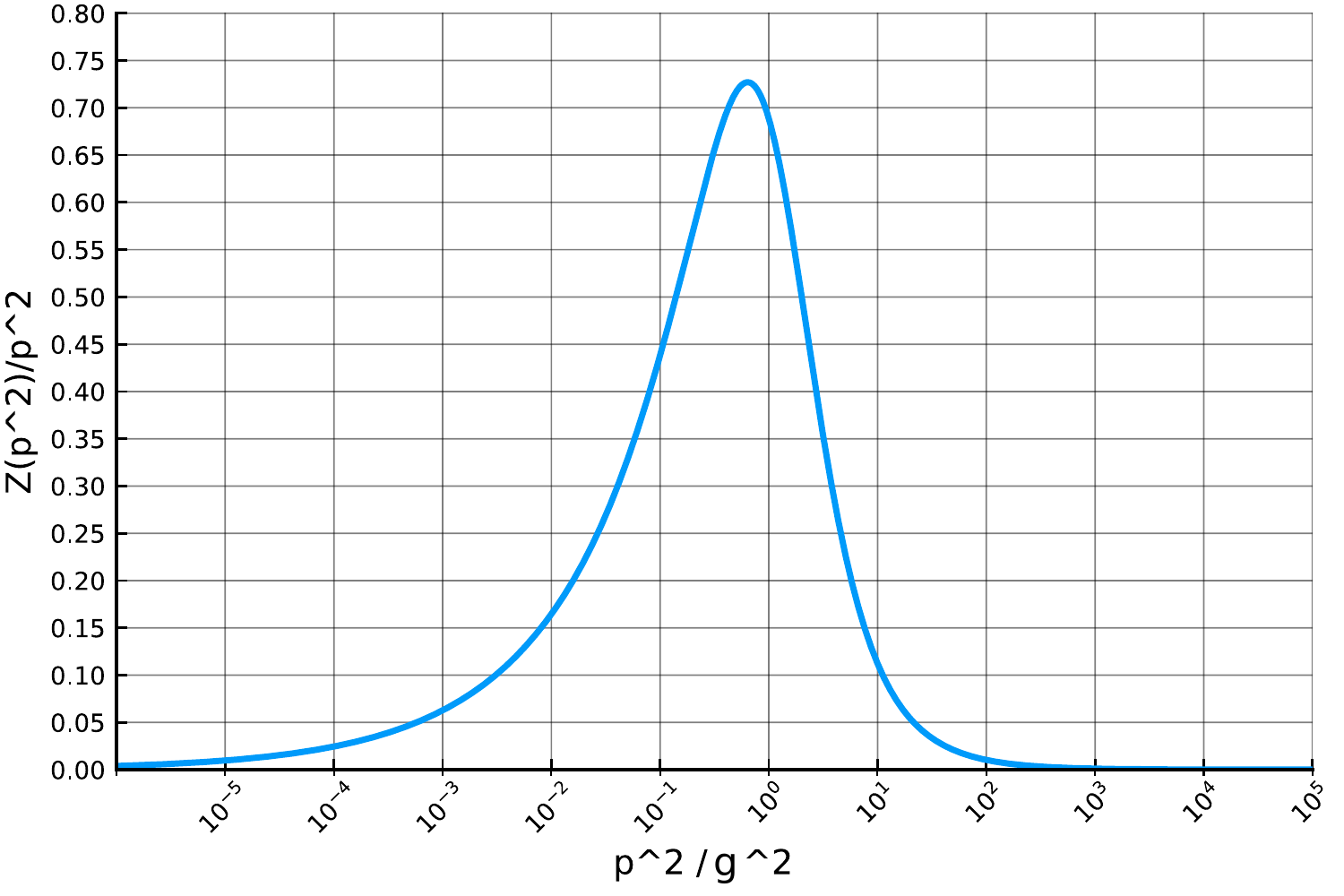}}{
				\caption{Gluon propagator.\myhfill}
				\label{fig:props_2d}
			}
			\ffigbox{\includegraphics[width=0.49\textwidth]{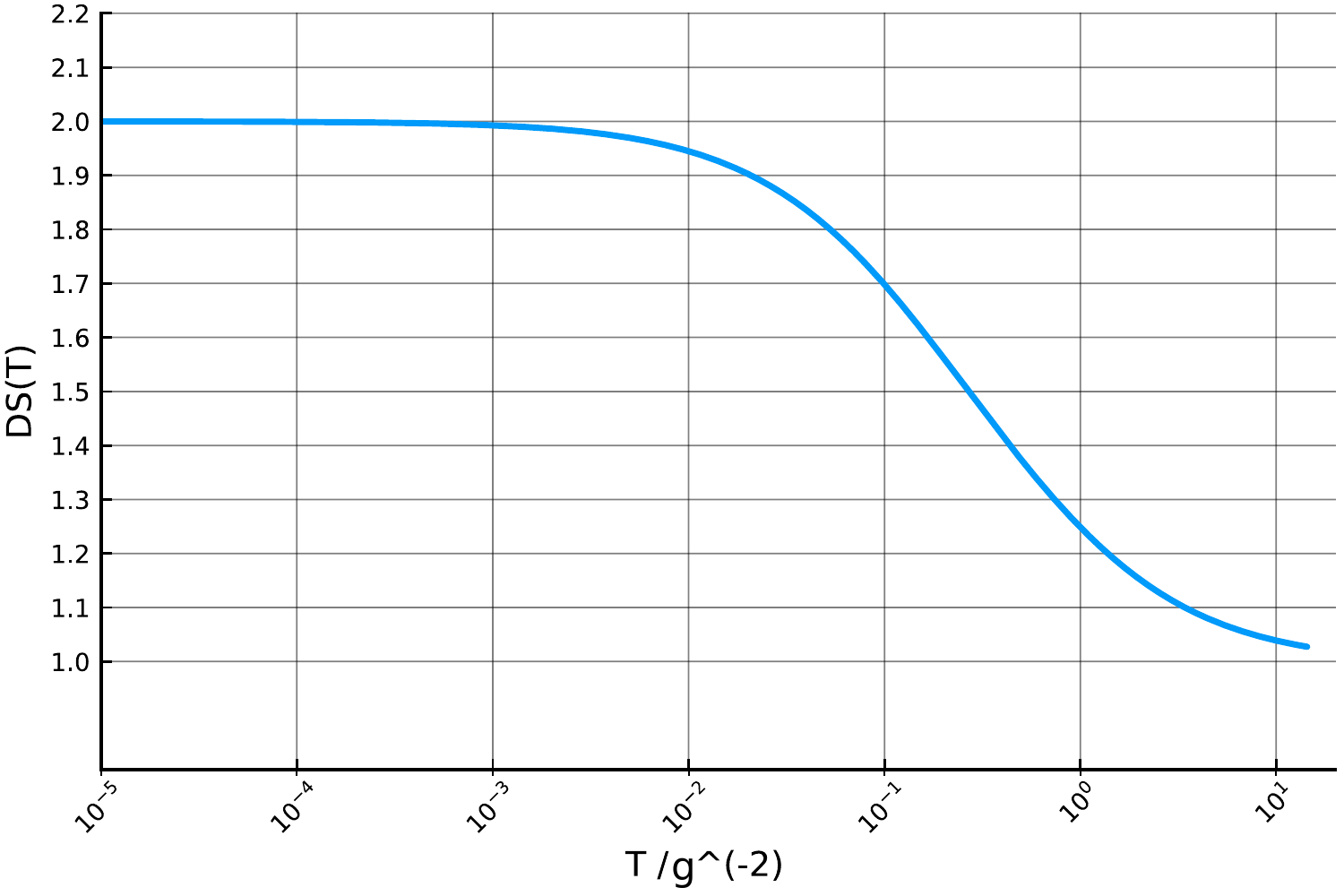}}{
				\caption{Spectral dimension.\myhfill}
				\label{fig:DS_2d}
			} 
		\end{subfloatrow}
	}{
		\caption{Gluon propagator and its spectral dimension for two-dimensional Yang-Mills theory.
		}
		\label{fig:2d}
	}
\end{figure*}

\begin{figure*}
	\ffigbox{
		\begin{subfloatrow}
			\ffigbox{\includegraphics[width=0.49\textwidth]{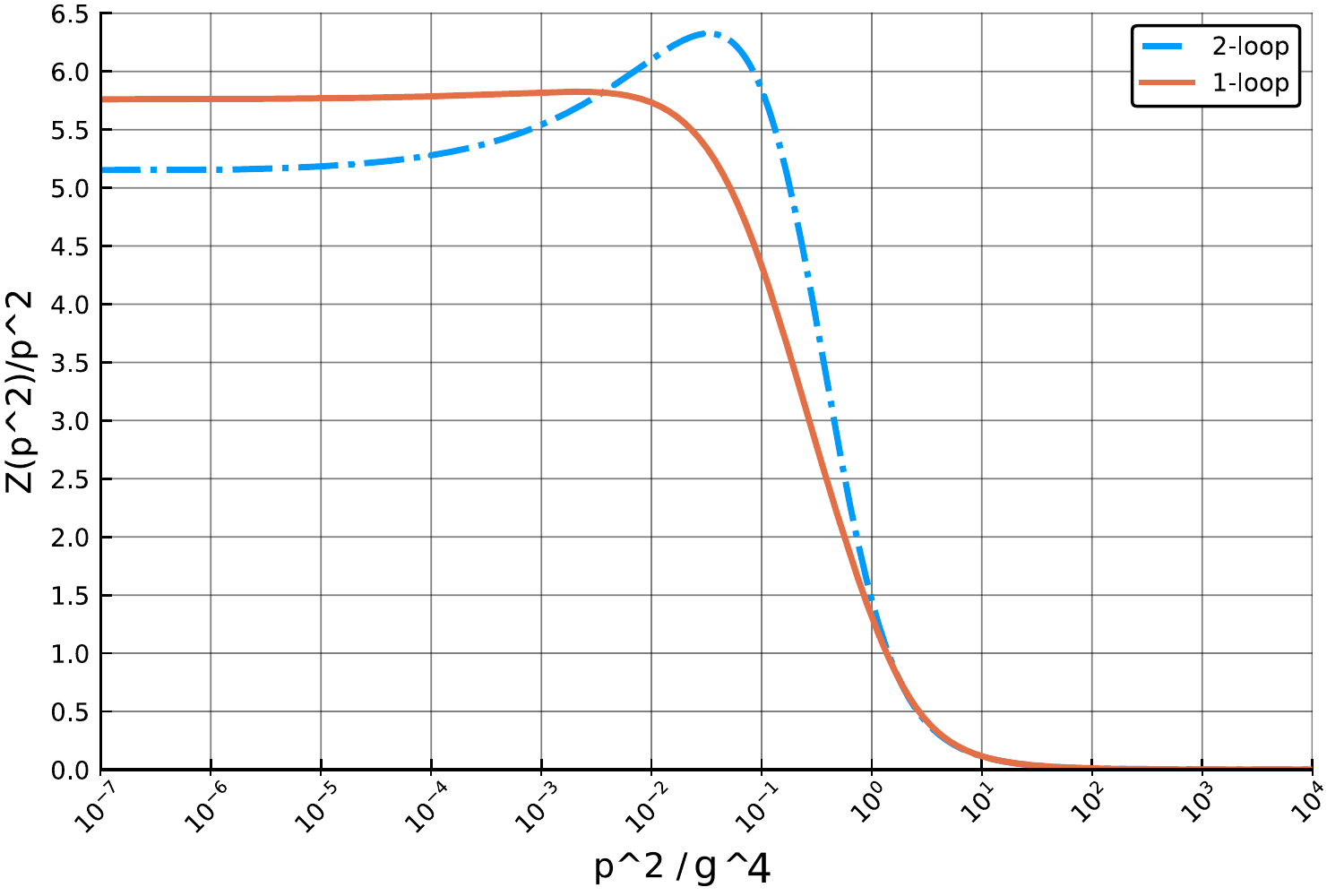}}{
				\caption{Gluon propagators.\myhfill}
				\label{fig:props_3d}
			}
			\ffigbox{\includegraphics[width=0.49\textwidth]{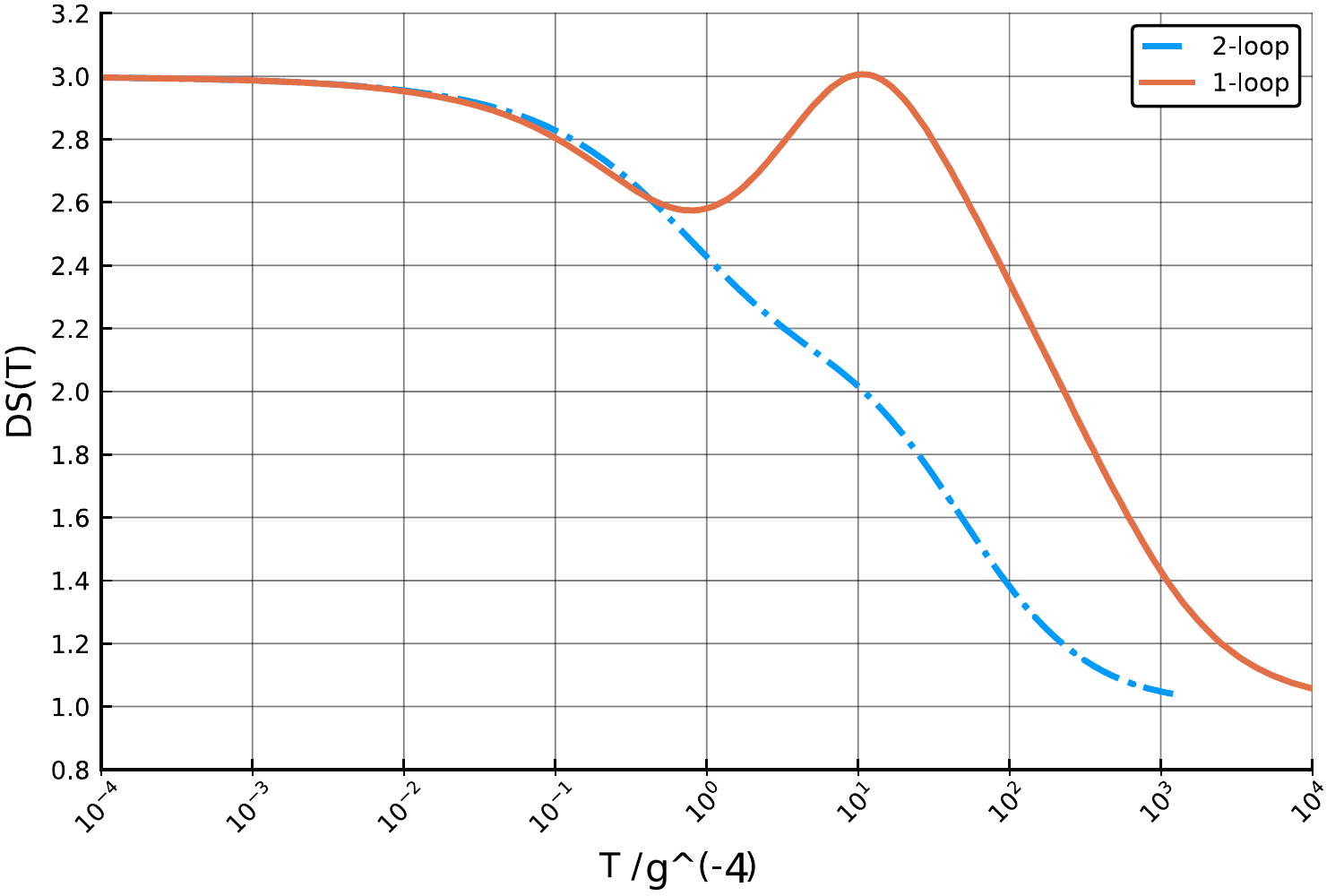}}{
				\caption{Spectral dimensions.\myhfill}
				\label{fig:DS_3d}
			} 
		\end{subfloatrow}
	}{
		\caption{
		Gluon propagators and their spectral dimensions for three-dimensional Yang-Mills theory obtained from two different truncations.
		}
		\label{fig:3d}
	}
\end{figure*}

\begin{figure*}
	\ffigbox{
		\begin{subfloatrow}
			\ffigbox{\includegraphics[width=0.49\textwidth]{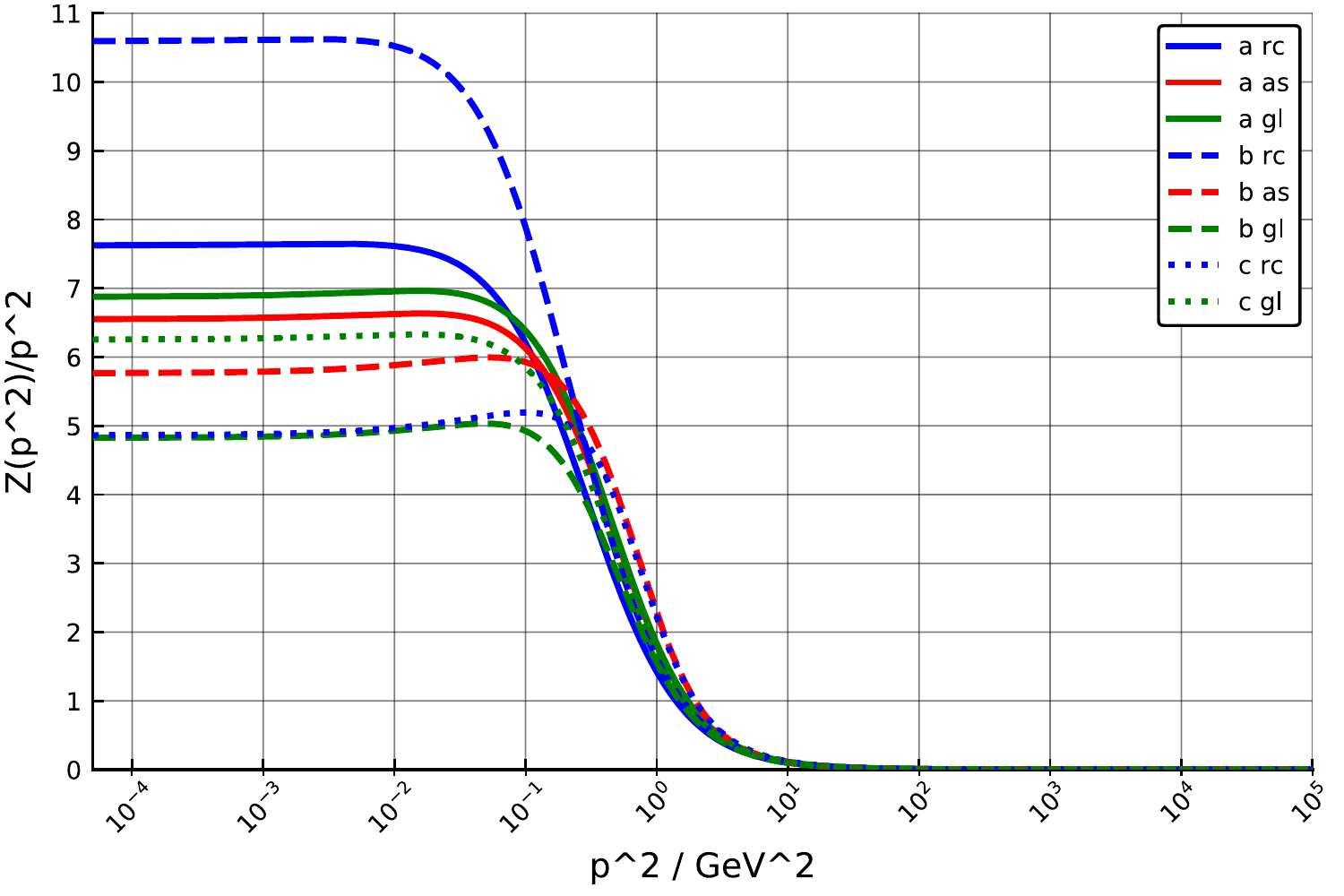}}{
				\caption{Gluon propagators.\myhfill}
				\label{fig:props_4d_1LDec}
			}
			\ffigbox{\includegraphics[width=0.49\textwidth]{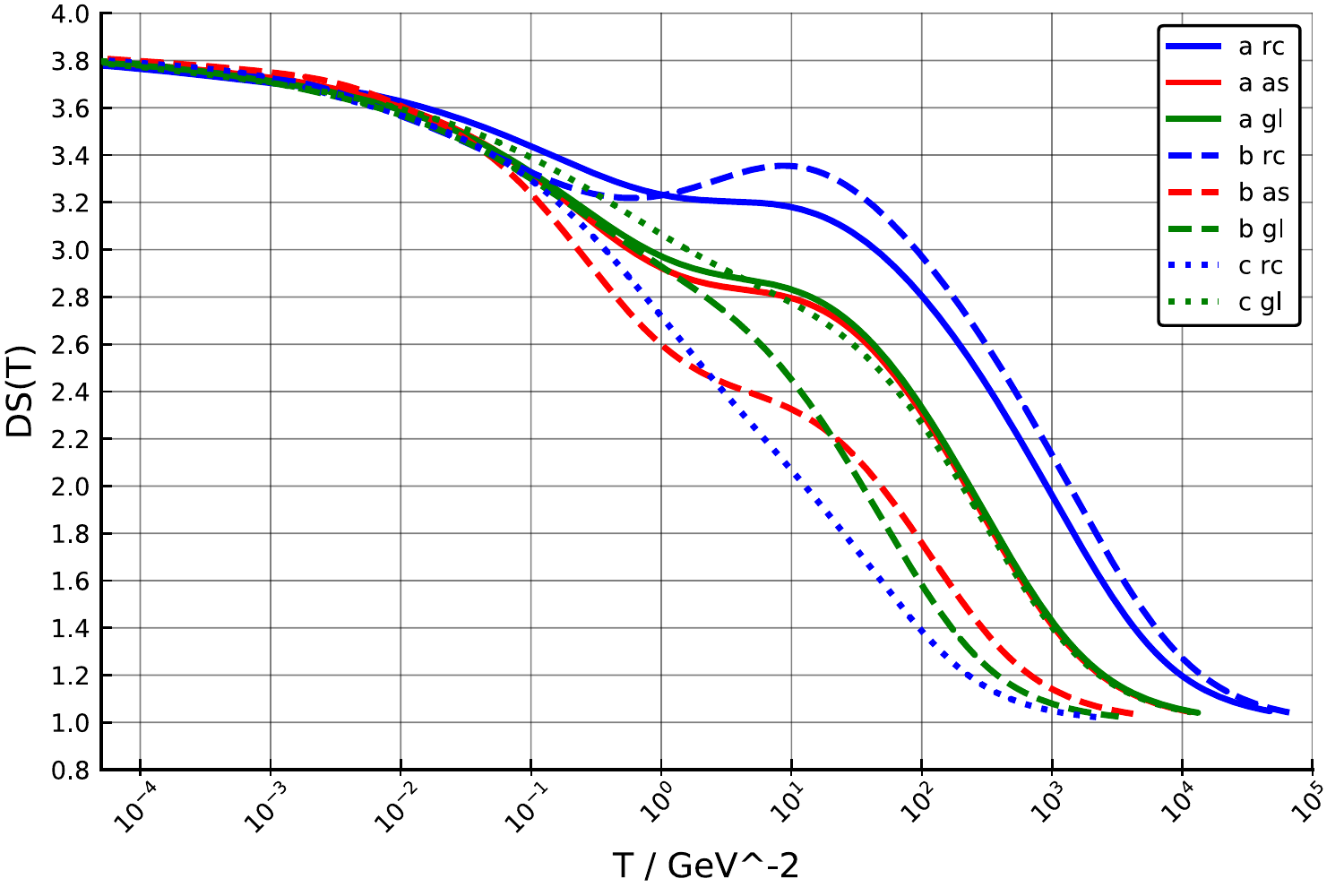}}{
				\caption{Spectral dimensions.\myhfill}
				\label{fig:DS_4d_1LDec}
			} 
		\end{subfloatrow}
	}{
		\caption{
		Gluon propagators and their spectral dimensions for four-dimensional Yang-Mills theory obtained from one-loop truncations of the decoupling type.
		The labels ``a'', ``b'' and ``c'' indicate three different models for the three-gluon vertex and ``rc'', ``as'' and ``gl'' three different subtraction methods for spurious divergences.
		}
		\label{fig:4d_1LDec}
	}
\end{figure*}

\begin{figure*}
	\ffigbox{
		\begin{subfloatrow}
			\ffigbox{\includegraphics[width=0.49\textwidth]{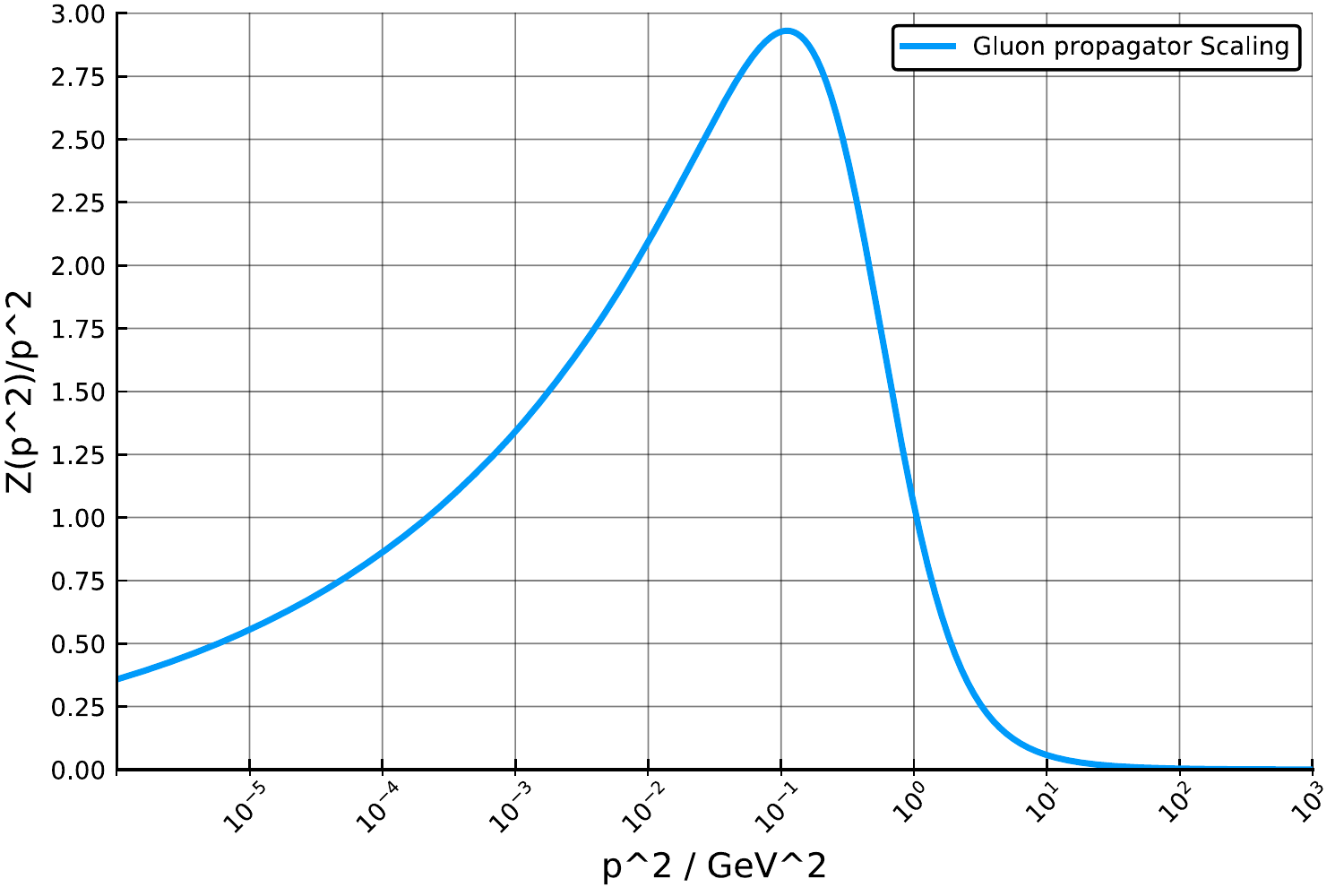}}{
				\caption{Gluon propagator.\myhfill}
				\label{fig:props_4d_1LSca}
			}
			\ffigbox{\includegraphics[width=0.49\textwidth]{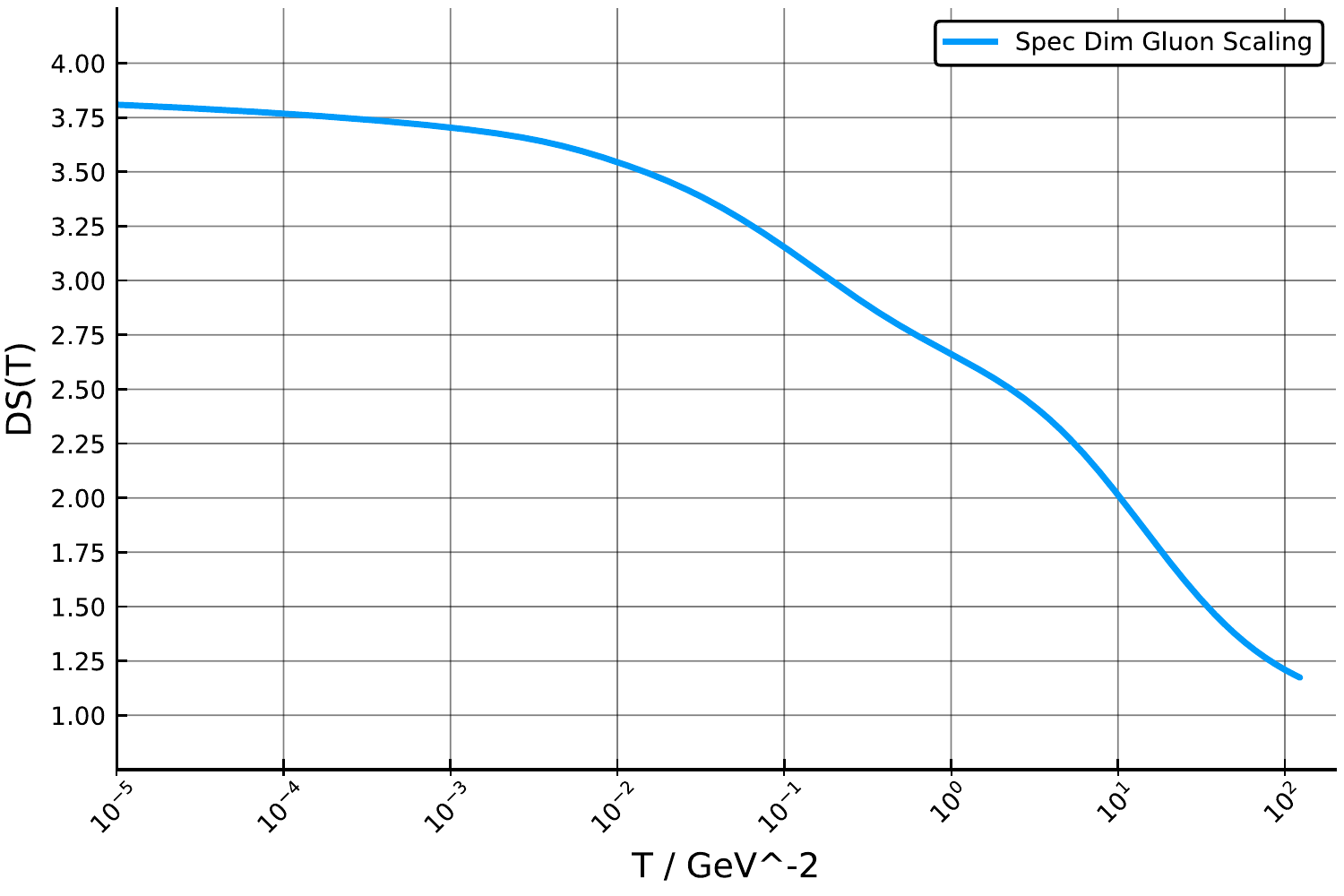}}{
				\caption{Spectral dimension.\myhfill}
				\label{fig:DS_4d_1LSca}
			} 
		\end{subfloatrow}
	}{
		\caption{
		Gluon propagator and its spectral dimension for four-dimensional Yang-Mills theory obtained from a one-loop truncation of the scaling type.
		}
		\label{fig:4d_1LSca}
	}
\end{figure*}

\subsection{Two dimensions}
\label{sec:input_2d}

The gluon propagator for two dimensions was obtained from a calculation of the gluon and ghost propagators using a bare ghost-gluon vertex and a model for the three-gluon vertex.
The data corresponds to the green lines in Fig.~15 of Ref.~\cite{Huber:2012zj}.
The gluon propagator is reproduced here in \fref{fig:props_2d}.
In two dimensions the coupling constant $g$ is a dimensionful quantity.
In our calculations we use the numeric value $1$ and plot dimensionful quantities in units of $g$.

It should be noted that in three and four dimensions a family of solutions exist for the propagators \cite{Boucaud:2008ji,Fischer:2008uz,Alkofer:2008jy}.
The endpoint of this family with an IR vanishing gluon propagator and an IR divergent ghost dressing function is called scaling solution \cite{vonSmekal:1997is,vonSmekal:1997vx} where all dressings are characterized by power laws \cite{Alkofer:2004it,Alkofer:2008bs,Fischer:2009tn}.
This solution necessarily always has a maximum in the gluon propagator.
For the other solutions both the gluon propagator and the ghost dressing are IR finite, e.g., \cite{Cornwall:1981zr,Cucchieri:2007md,Cucchieri:2008fc,Sternbeck:2007ug,Bogolubsky:2009dc,Dudal:2008sp,Boucaud:2008ji,Aguilar:2008xm,Fischer:2008uz,Alkofer:2008jy,Cyrol:2016tym,Huber:2018ned,Aguilar:2019kxz}.
In two dimensions, however, only the scaling solution can be realized \cite{Dudal:2008xd,Huber:2012zj,Zwanziger:2012xg}.

\subsection{Three dimensions}

For three-dimensional Yang-Mills theory we used two different truncations of Dyson-Schwinger equations.
The first one consists of the one-loop truncated propagator equations using a bare ghost-gluon vertex and the following model for the tree-level dressing of the three-gluon vertex motivated by a similar model in four dimensions \cite{Huber:2012kd}:
\begin{align}
\label{eq:3g_3d}
C^{AAA}(x,y,z)&=C^{AAA}_\text{UV}(x,y,z)+C^{AAA}_\text{IR}(x,y,z),\\
C^{AAA}_\text{UV}(x,y,z)&=\frac{1}{\sqrt{G(\overline{p}^2)}}, \\
C^{AAA}_\text{IR}(x,y,z)&=-G(\overline{p}^2)^3 \left(\frac{\Lambda^2_\text{3g}}{x+\Lambda^2_\text{3g}}\frac{\Lambda^2_\text{3g}}{y+\Lambda^2_\text{3g}}\frac{\Lambda^2_\text{3g}}{z+\Lambda^2_\text{3g}}\right)^4,
\end{align}
with $\overline{p}=(x+y+z)/2$ and $\Lambda_\text{3g}^2=10\,g^4$.
The second truncation consists of the full system of primitively divergent correlation functions, viz., the propagators and the ghost-gluon, three-gluon, and four-gluon vertices.
Details can be found in Ref.~\cite{Huber:2016tvc}.

In three dimensions the coupling constant $g$ is a dimensionful quantity.
In our calculations we use the numeric value $1$.
Due to ambiguities to set the scale \cite{Huber:2016tvc} -- which do not matter here -- we continue with raw data, so all dimensionful quantities should be considered as given in terms of $g$.

The spurious divergences related to the breaking of gauge covariance, see Ref.~\cite{Huber:2014tva} and references therein, are subtracted numerically by determining the coefficients of the linear and logarithmic cutoff dependence \cite{Huber:2016tvc}.

The propagators for the two truncations are compared in \fref{fig:props_3d}.

\subsection{Four dimensions}

In four dimensions, we test a variety of different inputs which can be grouped into four types.
For the DSE results the scale was set by putting the gluon dressing maximum to $0.94\,\text{GeV}$.

\begin{itemize}
 \item Solutions of a one-loop truncated gluon propagator DSE of decoupling type (1LDec):\newline
 To test the effects of the three-gluon vertex and the method for the subtraction of spurious divergences \cite{Huber:2014tva} we use combinations of three different variants of each.
 One combination does not yield a solution, so we have eight different setups.
 The three-gluon vertex models are called ``a'', ``b'', and ``c'' and the subtraction methods as ``gl'', ``rc'', and ``as''.
 Details are explained in App.~\ref{sec:3g_spurDivs}.
 \item Solutions of a one-loop truncated gluon propagator DSE of scaling type (1LSca):\newline
 For testing a scaling type solution of the gluon propagator results from Fig.~4 (red line) in Ref.~\cite{Huber:2014tva} are used.
 The subtraction of quadratic divergences corresponds to ``as'' and the model for the three-gluon vertex is an extended version of ``a''.
 \item Solutions of a two-loop truncated gluon propagator DSE of decoupling type  (2LDec):\newline
 We compare two different two-loop truncations.
 One uses models for the propagators \cite{Huber:2017txg} and another one includes the vertices dynamically \cite{Huber:2019wxx,Huber:2019ip}.
 \item Refined Gribov-Zwanziger fits (RGZ):\newline
 The refined Gribov-Zwanziger framework, as an extension of the original work by Gribov and Zwanziger \cite{Gribov:1977wm,Zwanziger:1989mf,Zwanziger:1992qr} taking into account condensates \cite{Dudal:2008sp,Gracey:2010cg,Dudal:2011gd}, provides a closed form for the gluon propagator that can be used for fits to lattice data.
 Here we use the parameters from Ref.~\cite{Cucchieri:2011ig} and a modified version that introduces a maximum into the propagator.
 The form we employ is
 \begin{equation}
 D_\text{RGZ}(p^2)=C \frac{p^2+s}{p^4+u^2 p^2+t^2}.
 \end{equation}
 For the parameters we use the following values \cite{Cucchieri:2011ig}:
 \begin{align}
 \label{eq:prop_RGZ}
 C&=0.784, \quad &s&=2.508\,\text{GeV}^2, \\
 t&=0.720\,\text{GeV}^2, \quad &u&=0.768\,\text{GeV}.
 \end{align}
 This gluon propagator does not exhibit a maximum, but we can modify the parameters to introduce one.
 Specifically, we use $u=0.15\,\text{GeV}$ for $D_\text{RGZ,max}(p^2)$.
\end{itemize}

\section{Results}
\label{sec:results}

The spectral dimension is calculated numerically from the second expression in \eref{eq:SpecDimDef}.

The results for two and three dimensions are shown in Figs.~\ref{fig:DS_2d} and \ref{fig:DS_3d}, respectively.
It clearly approaches the number of spacetime for short and one for long diffusion times as expected from the discussion in Sec.~\ref{sec:spec_dim}.
In two dimensions the transition is particularly smooth, which is most likely due to the distinct maximum in the propagator. 
For three dimensions, we can also compare different truncations and we find that indeed the behavior in the intermediate regime varies strongly:
The two-loop truncation shows a distinct maximum while the one-loop truncation is monotonic and has almost no structure.
This is reflected in an increase of the spectral dimension for intermediate diffusion times for the one-loop truncation.

In four dimensions we have a broader variety of setups.
All calculations (with the exception of the RGZ propagator) show that for long diffusion times the asymptotic value of one is approached.
For short diffusion times, we see that the spectral dimension approaches four only very slowly as expected from the discussion of logarithmic corrections in Sec.~\ref{sec:spec_dim}.
Details of that are discussed below for each case.
An exception are the RGZ propagators which have no logarithmic corrections and thus approach four directly.

We start with the details of the one-loop truncations shown in Figs.~\ref{fig:4d_1LDec} and \ref{fig:4d_1LSca}.
The scaling case is almost featureless.
However, comparing the different solutions for the decoupling case, we see that this should be attributed to the chosen three-gluon vertex model and the chosen subtraction scheme.
The setup for the scaling solution corresponds basically to ``b as'' with the exception that the three-gluon vertex model includes a zero crossing.
Changing the three-gluon vertex and/or the subtraction scheme, this changes and a plateau or even a maximum can appear.

\begin{figure}
\includegraphics[width=\textwidth]{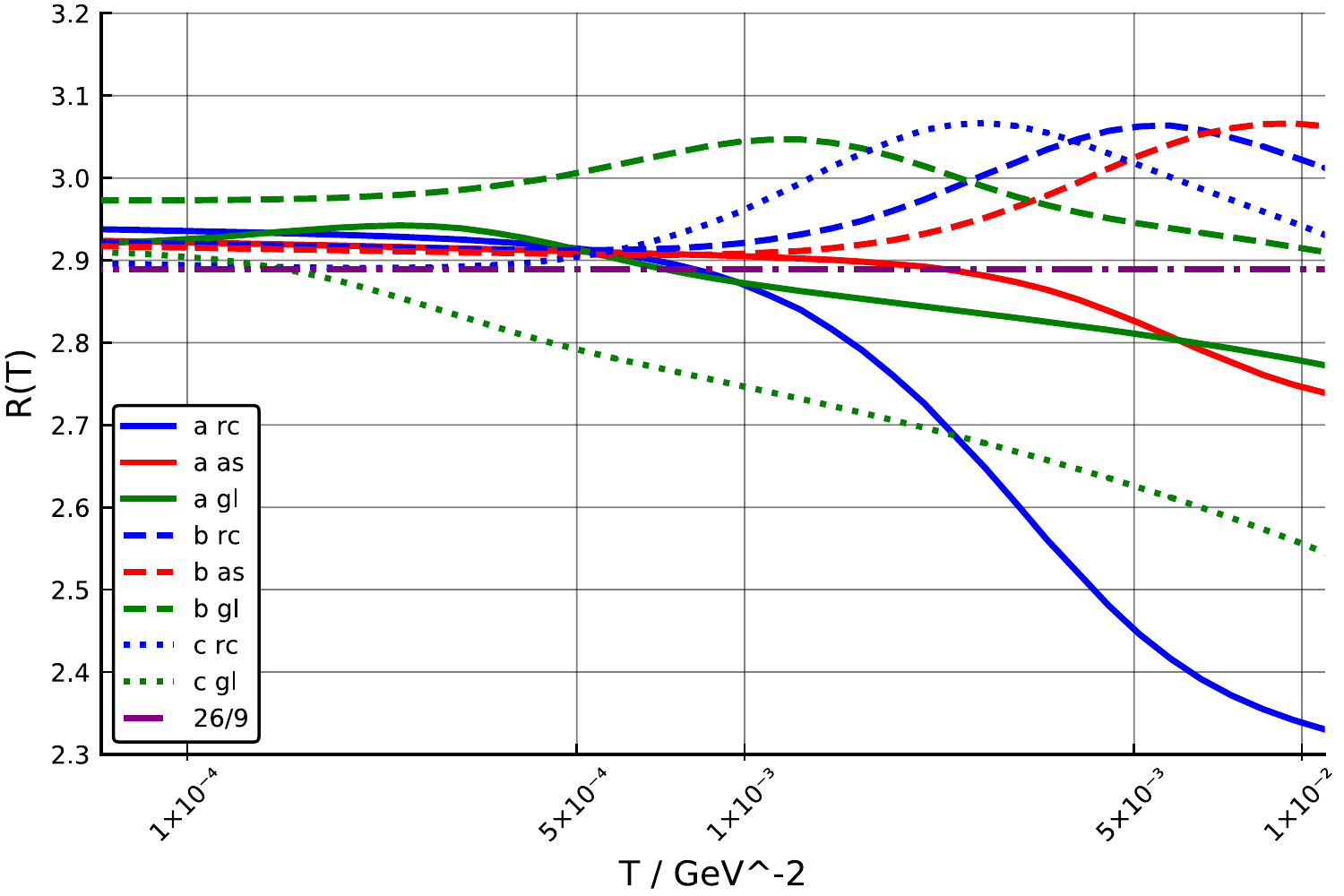} 
\caption{The ratio $R(T)$ for the one-loop truncated decoupling propagators from \fref{fig:props_4d_1LDec}.}
\label{fig:DS_4d_1L_quot}
\end{figure}

In Sec.~\ref{sec:spec_dim} we discussed that the ratio of the deviations of the spectral dimensions from four for different anomalous dimensions is equal to the ratio of anomalous dimensions.
We can test that by using the spectral dimension of the ghost propagator for comparison.
However, the ghost propagator is negative and the integral \eref{eq:SpecDimDef} does not converge.
We thus define a pseudospectral dimension $\widetilde{D}_S(T)$ that uses the negative ghost propagator, plot the quantity
\begin{align}
 R(T)=(4-D_S(T))/(4-\widetilde{D}_S(T))
\end{align}
and compare it to the ratio $\gamma/\delta$ of the one-loop anomalous dimensions of the gluon and ghost propagators, $\gamma=-13/22$ and $\delta=-9/44$, respectively.
This is shown in \fref{fig:DS_4d_1L_quot}.
For short diffusion times, the ratios are indeed similar, with the degree of coincidence depending on the truncation.

\begin{figure*}
	\ffigbox{
		\begin{subfloatrow}
			\ffigbox{\includegraphics[width=0.49\textwidth]{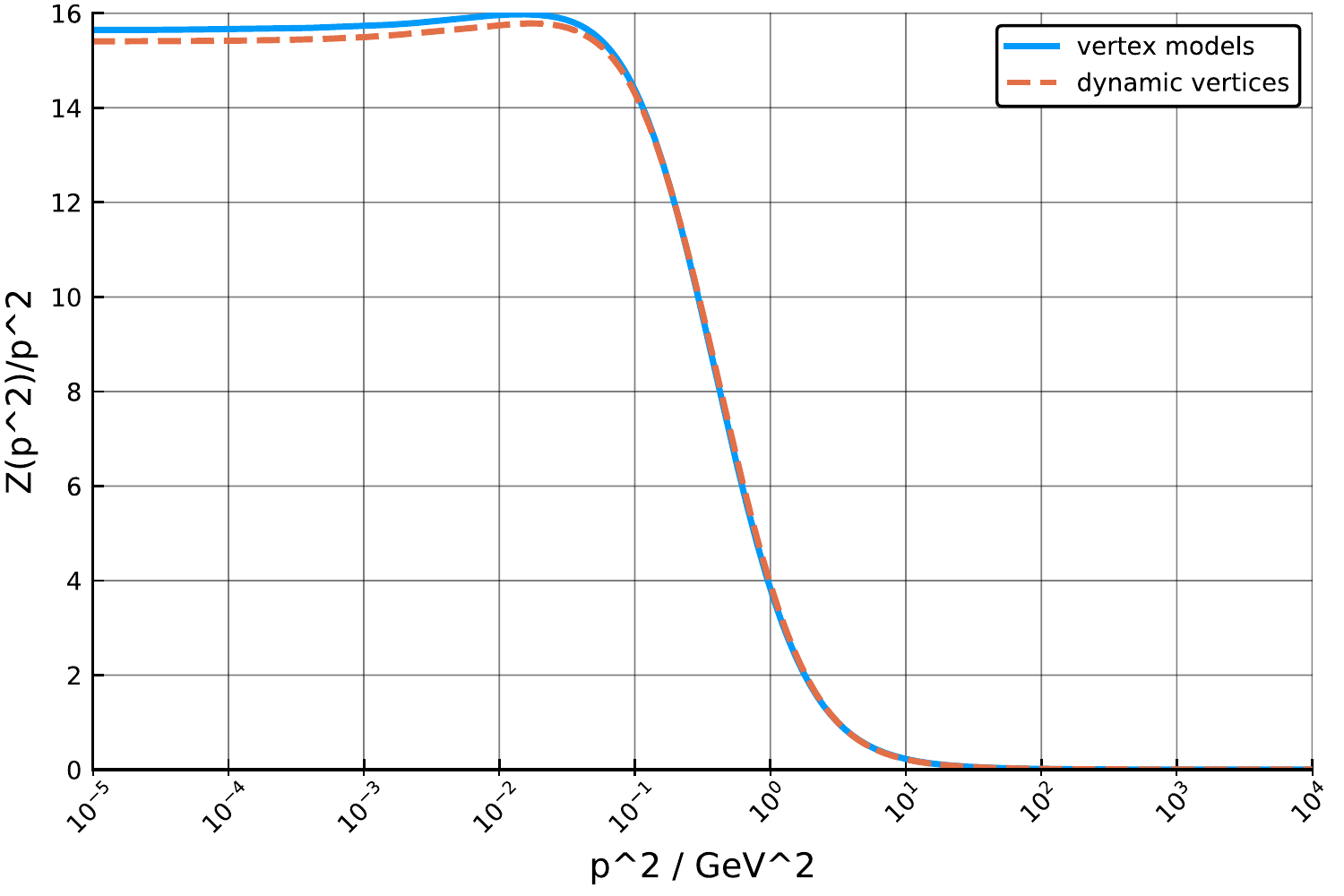}}{
				\caption{Gluon propagators.\myhfill}
				\label{fig:props_4d_2L}
			}
			\ffigbox{\includegraphics[width=0.49\textwidth]{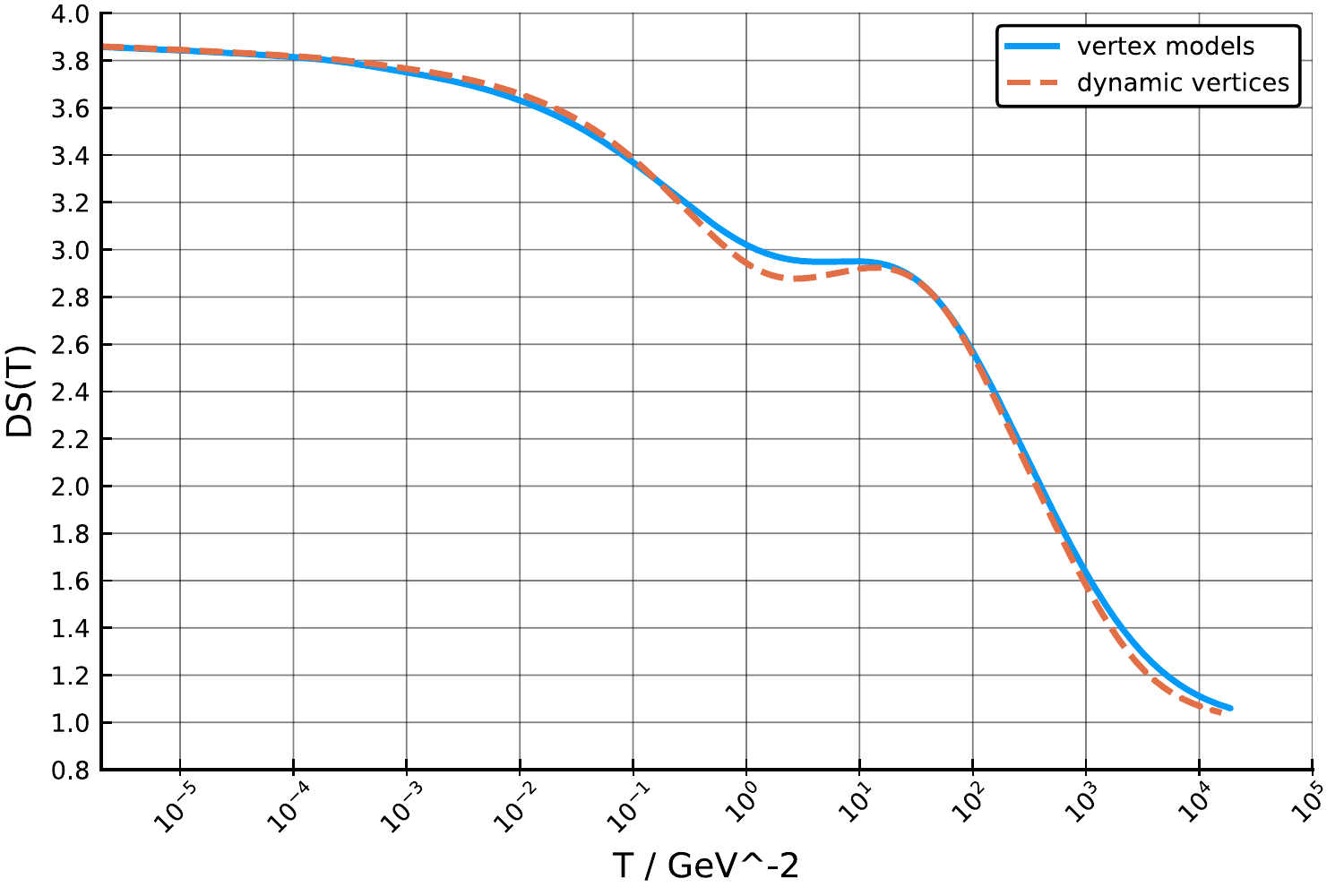}}{
				\caption{Spectral dimensions.\myhfill}
				\label{fig:DS_4d_2L}
			} 
		\end{subfloatrow}
	}{
		\caption{
		Gluon propagators and their spectral dimensions for four-dimensional Yang-Mills theory obtained from two-loop truncations using either models for the vertices \cite{Huber:2017txg} or dynamically calculated vertices \cite{Huber:2019wxx,Huber:2019ip}.
		}
		\label{fig:4d_2L}
	}
\end{figure*}

\begin{figure}
\centering
\includegraphics[width=\textwidth]{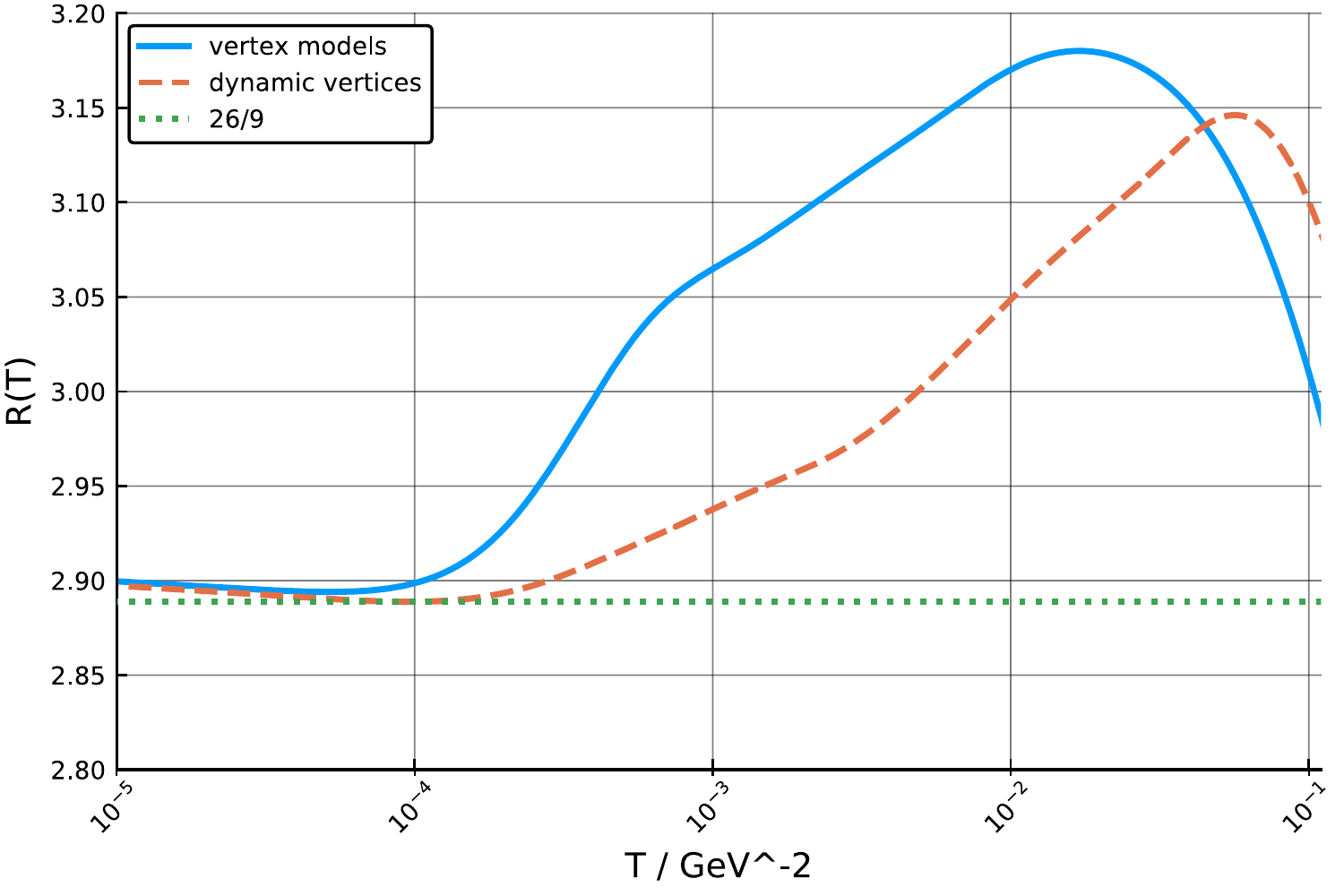} 
\caption{The ratio $R(T)$ for the two-loop truncated propagators from \fref{fig:props_4d_2L}.}
\label{fig:DS_4d_2L_quot}
\end{figure}

We turn now to the two-loop truncations.
The results for the spectral dimensions are shown in \fref{fig:DS_4d_2L}.
Both results show a similar behavior with a plateau in the intermediate regime.
An improvement compared to the one-loop truncations can be seen for the ratio of anomalous dimensions, shown in \fref{fig:DS_4d_2L_quot}, which approach the expected value very nicely to within $1\%$ for short diffusion times.
This could be related to the way the anomalous dimension is realized in these truncations.
For the one-loop truncations a renormalization group improvement term needs to be included in the three-gluon vertex model \cite{vonSmekal:1997vx,Fischer:2002eq,Huber:2012kd}, but for the two-loop truncations the anomalous dimension emerges correctly automatically \cite{Huber:2018ned,Huber:2019ip}.
Effects of such improvement terms were seen previously \cite{Huber:2014tva}  and also here by comparing different vertex models for the one-loop truncations.

\begin{figure*}
	\ffigbox{
		\begin{subfloatrow}
			\ffigbox{\includegraphics[width=0.49\textwidth]{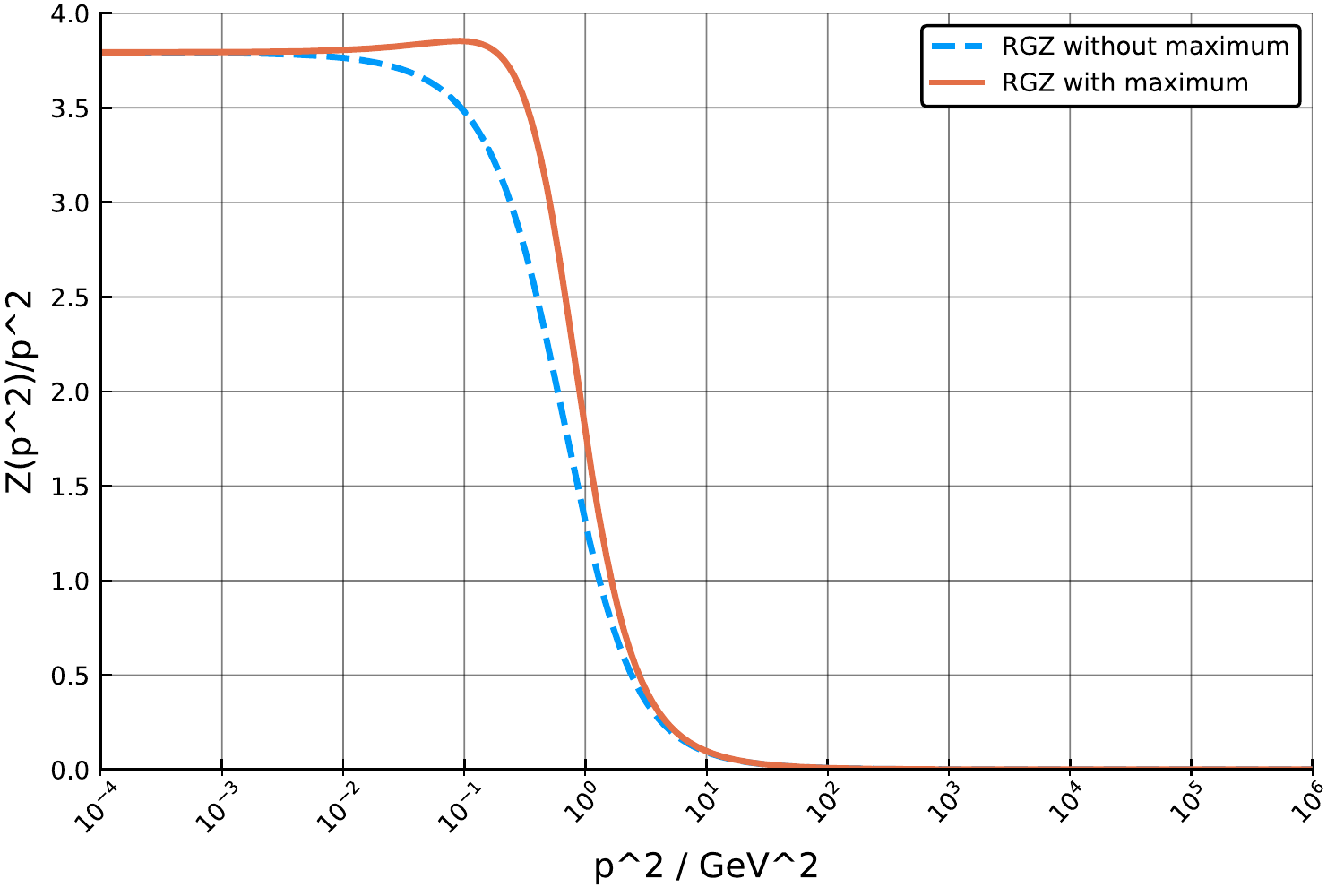}}{
				\caption{Gluon propagators.\myhfill}
				\label{fig:props_4d_RGZ}
			}
			\ffigbox{\includegraphics[width=0.49\textwidth]{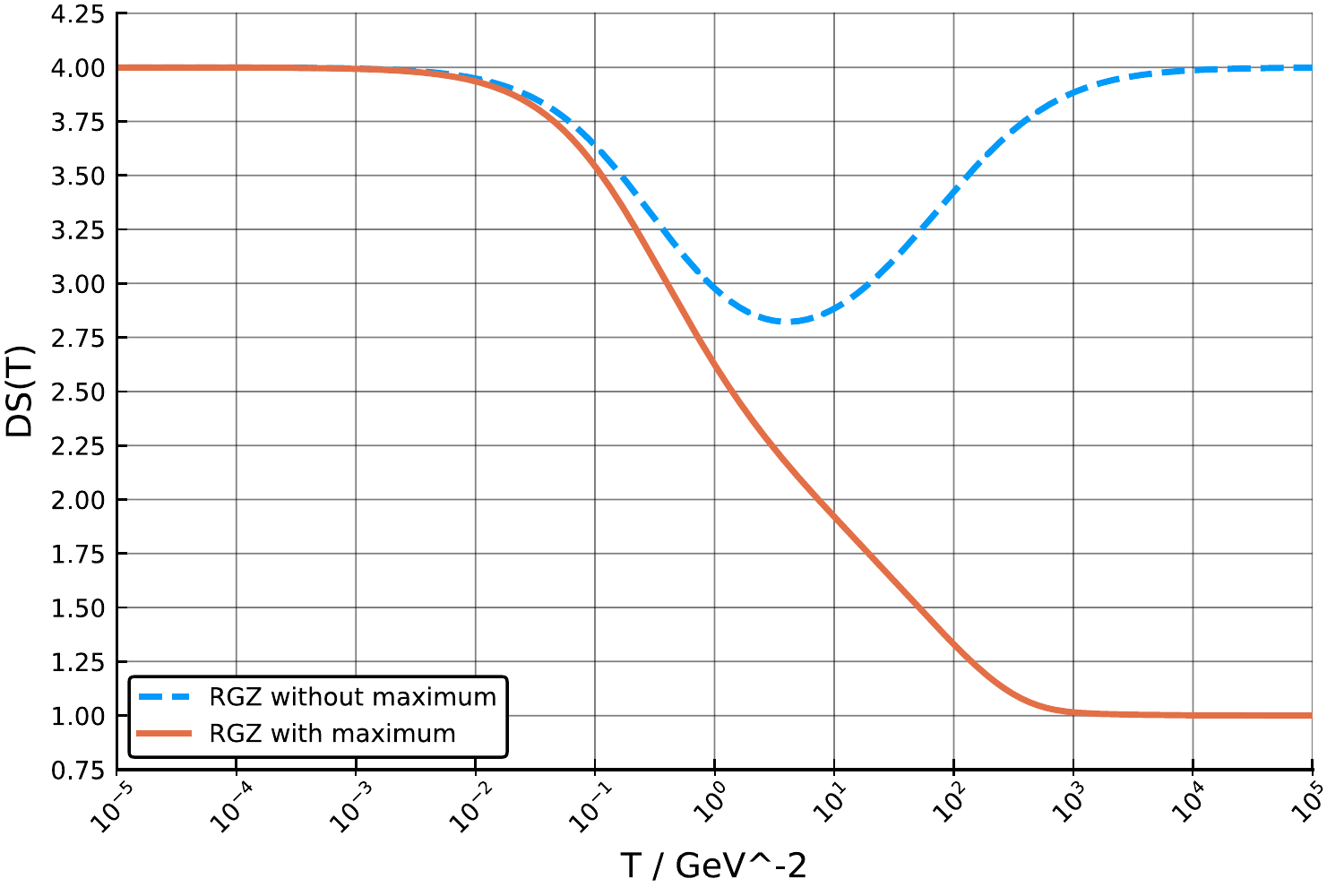}}{
				\caption{Spectral dimensions.\myhfill}
				\label{fig:DS_4d_RGZ}
			} 
		\end{subfloatrow}
	}{
		\caption{
		Gluon propagators and their spectral dimensions for four-dimensional Yang-Mills theory from the refined Gribov-Zwanziger framework \cite{Cucchieri:2011ig} and a modified version of it.
		}
		\label{fig:4d_RGZ}
	}
\end{figure*}

As last example we consider the RGZ model.
Since no logarithmic UV corrections are contained, the spectral dimension approaches four for short diffusion times very rapidly.
The spectral dimensions for the two propagators with and without maximum are shown in \fref{fig:DS_4d_RGZ}.
For large diffusion times, we see large differences between the two models which corroborate our discussion about the effect of a maximum in the propagator.
The standard RGZ propagator, which has no maximum and (after subtraction) a leading term proportional to $p^2$, turns back up for large diffusion times and approaches four, the canonical dimension of spacetime.
However, if we deform the propagator to contain a maximum, the asymptotic value for long diffusion times is one as expected from the general discussion in Sec.~\ref{sec:spec_dim}.

One general observation one can make from all these results is that the distinctiveness of the maximum in the gluon propagator is connected to the steepness of the spectral dimension in the intermediate regime.
All the propagators discussed have a maximum, except for the original RGZ propagator.
However, in some cases it is not very distinct, e.g., for the one-loop truncation in three dimensions or the setup ``c rc'' in four dimensions, see Figs.~\ref{fig:3d} and \ref{fig:4d_1LDec}, respectively.
In these cases, the spectral dimension goes up again before it approaches the asymptotic value.
In the three-dimensional example, it even goes up as far as three.
Given the analysis in Sec.~\ref{sec:spec_dim}, we know that there must be a jump in the asymptotic behavior of the spectral dimension once the maximum reaches zero.
This explains why for propagators with a flat maximum the resulting value for  the spectral dimension increases in between before it decreases again and approaches eventually its lowest and asymptotic value.

\section{Conclusions}
\label{sec:conclusions}

We have calculated the spectral dimensions for Yang-Mills gluon propagators as they are obtained from different truncations of Dyson-Schwinger equations or from the refined Gribov-Zwanziger framework.
The asymptotic behavior for short diffusion times reflects the spacetime dimension, but in four dimensions deviations from it are found which are a consequence of the propagator's perturbative (one-loop) anomalous dimension.

While short diffusion times are related to high energies, long diffusion times are not necessarily related to low energies.
We showed that for long diffusion times the result depends on the existence of a maximum of the propagator: The spectral dimension becomes then one.
Interestingly, this type of propagators is known to violate positivity.
While a spectral dimension of one for high diffusion times is not a necessary condition for positivity violation, it nevertheless establishes an interesting link to confinement.

The behavior of the spectral dimension for intermediate diffusion times mainly depends on the distinctiveness of the maximum in the gluon propagator.
Unfortunately, we could not identify dependencies on other aspects like the level of sophistication of the truncation with which the gluon propagator was calculated.
It is interesting to note that the relation of the maximum with the spectral dimension washed out any distinction between the decoupling and scaling type of solutions.
Largest distances, which clearly map to the longest diffusion time, do not relate to the deep infrared but to the momentum regime where the maximum is located.
Thus, one may speculate that confinement is not reflected in the infrared properties of correlation functions but in their properties somewhat below one GeV. This would then have some analogy to the phenomenon of dynamical chiral symmetry breaking, which in QCD is also driven by the interactions at the sub-GeV scale and not by the deep infrared.

While the calculation of the spectral dimension as a function of the diffusion time was done numerically, the main results we presented were obtained analytically.
Thus, they can be transferred directly to other cases, in particular to unquenched propagators and other gauges like linear covariant, Coulomb or maximally Abelian ones, for which results are available, e.g., \cite{Fischer:2003rp,Huber:2009wh,Alkofer:2011di,Huber:2015ria,Cyrol:2017ewj,Reinhardt:2017pyr,Aguilar:2012rz}, thereby extending the here presented study in order to elucidate the (in-)dependence of the results on the chosen gauge.

\vspace{10mm}

\section*{Acknowledgments}

Funding by the FWF (Austrian Science Fund) under Contract No. P 27380-N27 and the DFG (German Research Foundation) under Contract No. Fi970/11-1 are gratefully acknowledged.

\vspace{10mm}

\appendix

\section{Details on propagator solutions: One-loop truncated gluon propagator DSE with different models and subtraction methods (1LDec)}
\label{sec:3g_spurDivs}

Eight different combinations of three-gluon vertex models and subtraction methods are used.
The vertex models are the following ones:
\begin{widetext}
\begin{align}
C_\text{a}^{AAA}(p,q,-p-q)&=\frac{1}{Z_1}G\left(\frac{x+y+z+\Lambda_\text{IR}^2}{2}\right)^{2\alpha}Z\left(\frac{x+y+z+\Lambda_\text{IR}^2}{2}\right)^{2\beta},\\
C_\text{b}^{AAA}(p,q,-p-q)&=\frac{1}{Z_1}\left(\frac{G(\overline{p}^2)}{Z(\overline{p}^2)}\frac{\overline{p}^2}{\overline{p}^2+\Lambda_s^2}\right)^2,\\
C_\text{c}^{AAA}(p,q,-p-q)&=\frac{1}{Z_1}\frac{(G(y+\Lambda_\text{IR}^2)G(z+\Lambda_\text{IR}^2))^{1-a/\delta-2a}}{(Z(y+\Lambda_\text{IR}^2)Z(z+\Lambda_\text{IR}^2))^{1+a}},
\end{align}
\end{widetext}
with $x=p^2$, $y=q^2$, $z=(p+q)^2$, and $\overline{p}^2=(x+y+z)/2$.
$G$ and $Z$ are the ghost and gluon dressing functions, $\delta$ is the anomalous dimension of the ghost propagator and $Z_1$ the renormalization constant of the three-gluon vertex.
The models include terms for the renormalization group improvement necessary for the one-loop truncated gluon propagator DSE \cite{vonSmekal:1997vx,Fischer:2002hn,Huber:2012kd,Huber:2018ned}.
These models do not exhibit a zero crossing as seen with different methods, e.g., \cite{Pelaez:2013cpa,Aguilar:2013vaa,Blum:2014gna,Huber:2014bba,Eichmann:2014xya,Vujinovic:2014fza,Alkofer:2014taa,Williams:2015cvx,Huber:2018ned,Duarte:2016ieu,Athenodorou:2016oyh,Sternbeck:2016ltn}, but they are suppressed in the IR.
The remaining parameters are $\alpha=-2-6\delta$, $\beta=-1-3\delta$, and $a=3\delta$.
$\Lambda_\text{IR}$ is a small scale to suppress the IR divergence and $\Lambda_s$ a scale of the order of $1\,\text{GeV}$ to tune the IR suppression.
The actual values are not important and vary in each calculation due to the a posteriori determination of the physical scale.
More on model ``a'' can be found in \cite{Huber:2012kd}, on model ``b'' in \cite{Huber:2017txg} and on model ``c'' in \cite{Fischer:2002hn}.

The three methods to subtract the spurious divergences in the gluon propagator DSE are:
\begin{itemize}
\item Analytic subtraction (as): The divergent part of the self-energy is subtracted analytically \cite{Huber:2014tva}.
\item Gluon loop subtraction (gl): The gluon loop is modified to subtract spurious divergences \cite{Fischer:2002eq}.
\item Second renormalization condition (rc): The spurious divergences are removed by applying a second renormalization condition \cite{Collins:2008re,Meyers:2014iwa,Huber:2017txg}.
\end{itemize}

\section{Mass subtraction for nonmonotonic propagators leads to a negative spectral dimension}
\label{sec:neg_spec_dim}

We consider an inverse propagator with a minimum and subtract its value at $p^2=0$ to perform the mass-term subtraction.
We take this modified inverse propagator as $F(p^2)$.
It is negative in some interval $(a,b)$.
The related return probability $\mathcal{P}_{a,b}(T)$ is positive because the integrand 
\begin{align}
\label{eq:ret_DProb_Integ}
p^2 e^{-F(p^2) T} = p^2 e^{|F(p^2)|T}=p^2 \left(e^{|F(p^2)|}\right)^T
\end{align}
is positive in the complete interval.
Additionally, the value of the integral is increasing with T because of $e^{|F(p^2)|}>1$.
The derivative of the return probability $\mathcal{P}'_{a,b}(T)$ is positive too.
The derivative in \eqref{eq:ret_DProb_Integ} leads to
\begin{align}
\mathcal{P}'_{a,b}(T)\propto \int \limits_a^b dp^2~ p^2 |F(p^2)|\left(e^{|F(p^2)|}\right)^T.
\end{align}
Again, all factors of the integrand are positive and so is the integral itself.
It also increases with $T$.

We call the contribution of the return probability coming from the positive part of $F(p^2)$ $\mathcal{P}_{rest}(T)$.
It is positive and decreases with $T$.
Its derivative $\mathcal{P}'_{rest}(T)$, though, is negative due to the factor $-F(p^2)$ in the integrand.
The spectral dimension reads
\begin{align}
D_S(T)=-2 \frac{\mathcal{P}'(T)T}{\mathcal{P}(T)}=-2\frac{(\mathcal{P}'_{a,b}(T)+\mathcal{P}'_{rest}(T))T}{\mathcal{P}_{a,b}(T)+\mathcal{P}_{rest}(T)}.
\end{align}
If $T$ is sufficiently large, $|\mathcal{P}'_{a,b}(T)|>|\mathcal{P}'_{rest}(T)|$.
Hence, the numerator is positive in this case.
The denominator, is also positive for any $T$.
Thus, the resulting spectral dimension is negative for large diffusion times.

\section{Asymptotic behavior of the spectral dimension}
\label{sec:asymptotics}

We give the proofs for the two cases for the asymptotic behavior of the spectral dimension considered in Sec.~\ref{sec:spec_dim}.
The first one is for a polynomial form of the inverse propagator.
This case was already considered in Ref.~\cite{Alkofer:2014raa} and we give the proof here for convenience of the reader.ß
We consider an inverse propagator of the form
\begin{align}
\label{eq:F_poly}
F(p^2)=(p^2)^{n_\text{max}}+\ldots+(p^2)^{n_\text{min}}.
\end{align}
For the case $F(p^2)=(p^2)^n$, the integral for the return probability, \eref{eq:return_prob}, can be done analytically.
First, we perform a change of variable $p\rightarrow (t/T)^{1/2n}$.
The return probability is then proportional to
\begin{align}
\mathcal{P}(T)\propto T^{-\frac{d}{2n}}\frac{1}{2n}\int_0^\infty dt\, t^{\frac{d}{2n}-1} e^{-t} = T^{-\frac{d}{2n}}\frac{1}{2n}\Gamma\left(\frac{d}{2n}\right),
\end{align}
from which the spectral dimension is calculated as
\begin{align}
 D_S(T)=-2 \frac{d\, \ln P(T)}{d\,\ln T}=\frac{d}{n}.
\end{align}
If the inverse propagator has the polynomiaßl form of \eref{eq:F_poly}, the power with the largest/smallest $n$ will dominate for short/long diffusion times and we obtain
\begin{align}
 \lim_{T\rightarrow 0}D_S(T)&=\frac{d}{n_\text{max}},\\
 \lim_{T\rightarrow \infty}D_S(T)&=\frac{d}{n_\text{min}}.
\end{align}

The second case is for the inverse propagator of the form $(p^2-p_0^2)^c$ with $c>0$.
Using \eref{eq:return_prob_k} with $F^{-1}(k^2)=(k^2)^\frac{1}{c}+p_0^2$ and $F'(p^2)=c\,(k^2)^{1-\frac{1}{c}}$, this leads to the following integral:
\begin{align}
 \mathcal{P}(T) \propto \int_0^\infty dk\,\frac{k^{\frac{2}{c}-1}\left( k^\frac{2}{c}+p_0^2 \right)^{\frac{d}{2}-1}}{c} e^{-k^2\,T}.
\end{align}
For $d=4$ this evaluates to
\begin{align}
 \mathcal{P}(T) \propto \frac{T^{-\frac{2}{c}}\left(p_0^2\, T^\frac{1}{c}\, \Gamma\left(\frac{1}{c}\right)+\Gamma\left(\frac{2}{c}\right)\right)}{2c}
\end{align}
so that the spectral dimension becomes
\begin{align}
 D_S(T)=\frac{2 \,p_0^2\, T^{\frac{1}{c}} \Gamma \left(\frac{1}{c}\right)+4\, \Gamma \left(\frac{2}{c}\right)}{c \,p_0^2\, T^{\frac{1}{c}} \Gamma \left(\frac{1}{c}\right)+c\, \Gamma \left(\frac{2}{c}\right)}.
\end{align}
In the limit of $T\rightarrow \infty$, this leads to $2/c$.
For two dimensions we obtain
\begin{align}
\mathcal{P}(T) \propto \frac{T^{-\frac{1}{c}}\Gamma\left(\frac{1}{c}\right)}{2c}
\end{align}
and also $2/c$ for $T\rightarrow \infty$.
We tested numerically that this is also true for three dimensions.

\vspace{10mm} 

~

\bibliographystyle{utphys_mod}
\bibliography{literature_spectral_dim}

\end{document}